\documentclass[aps,prl,reprint,groupedaddress]{revtex4-1}

\newcommand{\be}{\begin{equation}}
\newcommand{\ee}{\end{equation}}

\usepackage{verbatim}
\usepackage{latexsym,amssymb,float}
\usepackage{setspace}
\usepackage{graphicx}
\usepackage{epsfig}
\usepackage{amsmath}
\usepackage{bm}

\begin{document}

\title{Imaging the Spatial Form of a Superconducting Order Parameter via Josephson Scanning Tunneling Spectroscopy}
\author{Martin Graham}
\author{Dirk K. Morr}
\affiliation{University of Illinois at Chicago, Chicago, IL 60607, USA}

\date{\today}

\begin{abstract}
Motivated by recent experiments, we investigate Josephson scanning tunneling spectroscopy in an $s$-wave superconductor. We demonstrate that the spatial oscillations in the superconducting order parameter induced by defects can be spatially imaged through local measurements of the critical Josephson current, providing unprecedented insight into the nature of superconductivity. The spatial form of the Josephson current reflects the nature of the defects, and can be used to probe defect-induced phase transitions from an $S=0$ to an $S=1/2$ ground state.
\end{abstract}


\maketitle

Imaging the spatial variations of superconducting order parameters has been a long-sought goal, as it could provide direct insight into the nature of exotic superconducting phases ranging from the Fulde-Ferrell-Larkin-Ovchinnikov state \cite{Ful64,Lar65,Bia03,Mat07} in the presence of magnetic fields, and intrinsically disordered superconductors \cite{Lang02} to the pair-density wave state predicted to exist in the cuprate superconductors \cite{Chen04,Berg09,Lee14,Fra15,Wang15}. As the oscillations of the superconducting order parameter are expected to occur on the length scale of a few lattice constants, and their detection hence requires near atomic resolution, recent experimental efforts have focused on the development of Josephson scanning tunneling spectroscopy (JSTS) \cite{Sma01,Ham15,Jae16,Ran16}. The idea underlying JSTS is that the Josephson current, $I_J$, \cite{Jos62} flowing between a superconducting JSTS tip and a superconductor probes the order parameter of the latter \cite{Amb63}. Using this technique, Hamidian {\it et al.} \cite{Ham15} have argued that the spatial oscillations in $I_J$ induced by defects in the cuprate superconductor Bi$_2$Sr$_2$CaCu$_2$O$_{8+x}$ provide evidence for the existence of a pair-density wave. Complementary to this study, Randeria {\it et al.} \cite{Ran16} showed that pair-breaking magnetic Fe atoms located on the surface of the $s$-wave superconductor Pb lead to a suppression of the local Josephson current. So far, however, there has been no proof for the assumption that the experimentally measured spatial variations of the Josephson current indeed reflect those of the superconducting order parameter.

\begin{figure}
\includegraphics[width=8cm]{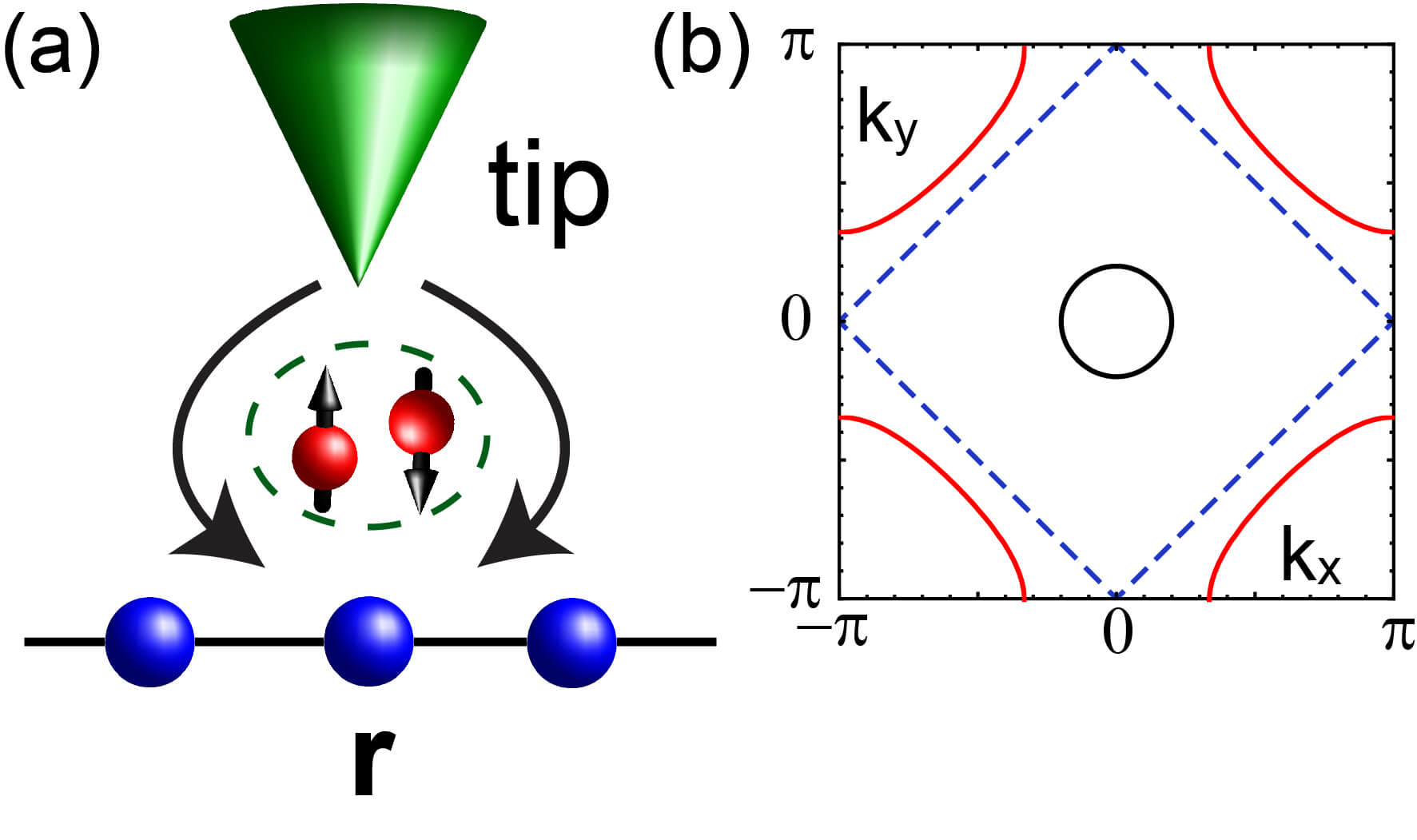}%
 \caption{Schematic representation of Cooper pair Josephson tunneling from a superconducting JSTS tip into an $s$-wave superconductor. (b) Fermi surfaces for three different electronic structures of the $s$-wave superconductor.}
 \label{fig:tunneling}
 \end{figure}

In this article, we provide this missing proof by theoretically demonstrating that even short length scale fluctuations of the superconducting order parameter can be spatially imaged through local measurements of the Josephson current, thus opening unprecedented possibilities for gaining insight into the nature of superconductivity. Using a Keldysh non-equilibrium Green's function formalism, we investigate the local Josephson current between a superconducting JSTS tip with $s$-wave symmetry, and an $s$-wave superconductor [as schematically shown in Fig.~\ref{fig:tunneling}(a)], and its relation to the local superconducting order parameter. We demonstrate that spatial oscillations in the superconducting order parameter, $\Delta({\bf r})$, induced by both magnetic and non-magnetic defects can be imaged at the atomic length scale by measuring the spatial form of the critical Josephson current, $I_c({\bf r})$. Moreover, for magnetic defects, the existence of defect-induced Shiba states \cite{Shi68,Yaz97,Sal97,Hat15} inside the superconducting gap can lead to an enhanced Josephson current in the vicinity of the defect, an effect which is absent for non-magnetic defects. Furthermore, we demonstrate that JSTS can be employed to detect phase transitions between different spin ground states of the superconductor, as occur in the presence of magnetic defects \cite{Sak70,Sal97,Morr06,Hat15}.  Finally, we show that JSTS can even image virtual defects which are created using quantum interference effects. These results demonstrate that JSTS provides unprecedented possibilities to gain insight into the spatial nature of superconducting order parameters.

Starting point for investigating the relation between the spatial form of the critical Josephson current and the superconducting order parameter around defects in an $s$-wave superconductor is the Hamiltonian $H=H_s + H_{tip} + H_{tun}$ where
\begin{align}
H_s &=-t \sum_{\langle {\bf r, r'} \rangle ,\sigma}  c^\dagger_{{\bf r} \sigma} c_{{\bf r^\prime} \sigma} -\mu \sum_{ {\bf r}, \sigma} c^\dagger_{{\bf r} \sigma} c_{{\bf r}\sigma} \nonumber \\
& -\sum_{\bf r} \left[  \Delta({\bf r}) c_{{\bf r}\uparrow}^{\dagger} c_{{\bf r}\downarrow}^\dagger + H.c. \right] \nonumber \\
& + \sum_{{\bf R},\alpha,\beta} (U_0 {\hat 1}_{\alpha\beta} +J_0 \sigma^z_{\alpha\beta} )c_{{\bf R},\alpha}^\dagger c_{{\bf R},\beta}
\label{eq:Hs}
\end{align}
Here, $-t$ is the electronic hopping between nearest-neighbor sites ${\bf r}$ and ${\bf r^\prime}$, $\mu$ is the chemical potential, and $c^\dagger_{{\bf r} \sigma}$ ($c_{{\bf r} \sigma}$) creates (annihilates) an electron with spin $\sigma$ at site ${\bf r}$. $\Delta({\bf r})$ is the superconducting order parameter with $s$-wave symmetry at site {\bf r} in the superconductor and $U_0$ and $J_0$ are the non-magnetic and magnetic scattering strengths of a defect located at site ${\bf R}$, with the last sum running over all defect sites. Unless otherwise noted, we set $\mu=-3.618t$, yielding the circular Fermi surface shown in Fig.~\ref{fig:tunneling}(b). In the presence of defects, we self-consistently compute the local superconducting order parameter in the superconductor using
\begin{align}
\Delta({\bf r})=-\frac{V_0}{\pi}\int_{-\infty}^\infty d\omega n_F(\omega) \text{Im}[F_s({\bf r},{\bf r},\omega)] \label{eq:OP}
\end{align}
where $V_0$ is the superconducting pairing potential, $n_F(\omega)$ is the Fermi distribution function, and $F_s({\bf r},{\bf r},\omega)$ is the local, retarded anomalous Green's function of the $s$-wave superconductor (see supplemental information (SI) Sec.~I). We model the JSTS tip as an atomically sharp site, described by the Hamiltonian $H_{tip}=H_{tip}^n+H_{tip}^{sc}$, where $H_{tip}^{n}$ represents the normal state electronic structure of the tip, and
\begin{align}
H_{tip}^{sc}= -\Delta_{tip} d^\dagger_\uparrow d^\dagger_\downarrow - \Delta_{tip} d_\downarrow d_\uparrow
\end{align}
its superconducting correlations. Here, $\Delta_{tip}$ is the superconducting $s$-wave gap in the tip, and $d^\dagger_\sigma$ ($d_\sigma$) creates (annihilates) an electron with spin $\sigma$ in the tip. Finally, the tunneling of electrons between the tip and a site ${\bf r}$ in the $s$-wave superconductor is described by
\begin{align}
H_{tun}=-t_0 \sum_\sigma(c_{{\bf r}, \sigma}^\dagger d_\sigma+d^\dagger_\sigma c_{{\bf r}, \sigma})
\end{align}

A DC Josephson current \cite{Jos62} arises from a phase difference between the superconducting order parameters of the tip and the $s$-wave superconductor, described by
\begin{equation}
\Delta({\bf r}) = |\Delta({\bf r})| e^{i \Phi_s} \qquad \Delta_{tip} = |\Delta_{tip}| e^{i \Phi_t}
\label{eq:Delta_s}
\end{equation}
This phase difference can be gauged away \cite{Cue96}, yielding a tunneling parameter that depends on the phase difference
\begin{align}
t_0 \rightarrow t_0 e^{i (\Phi_s-\Phi_t)/2} = t_0 e^{i \Delta \Phi/2} \ .
\end{align}
allowing us to take $\Delta({\bf r})$ and $\Delta_t$ as real parameters below. Using the Keldysh Green's function formalism \cite{Kel65,Ram86}, we then obtain that the DC-Josephson current between the tip and a site ${\bf r}$ in the s-wave superconductor to lowest order in the hopping $t_0$ is given by \cite{Cue96}
\begin{align}
I_{J}({\bf r}) &=8\frac{e}{\hbar} t_0^2  \sin{\left(\Delta \Phi\right)}  \int \frac{d\omega}{2\pi}n_F(\omega)\text{Im}[F_s({\bf r},{\bf r},\omega)F_t(\omega)] \nonumber \\
 & \equiv I_c({\bf r}) \sin{\left(\Delta \Phi\right)}
\label{eq:Ieq}
\end{align}
where $F_t$ is the retarded anomalous Green's function of the tip (see SI Sec.~I), $I_c$ is the critical Josephson current, and we set $T=0$ below.

\begin{figure}[t]
\includegraphics[width=8cm]{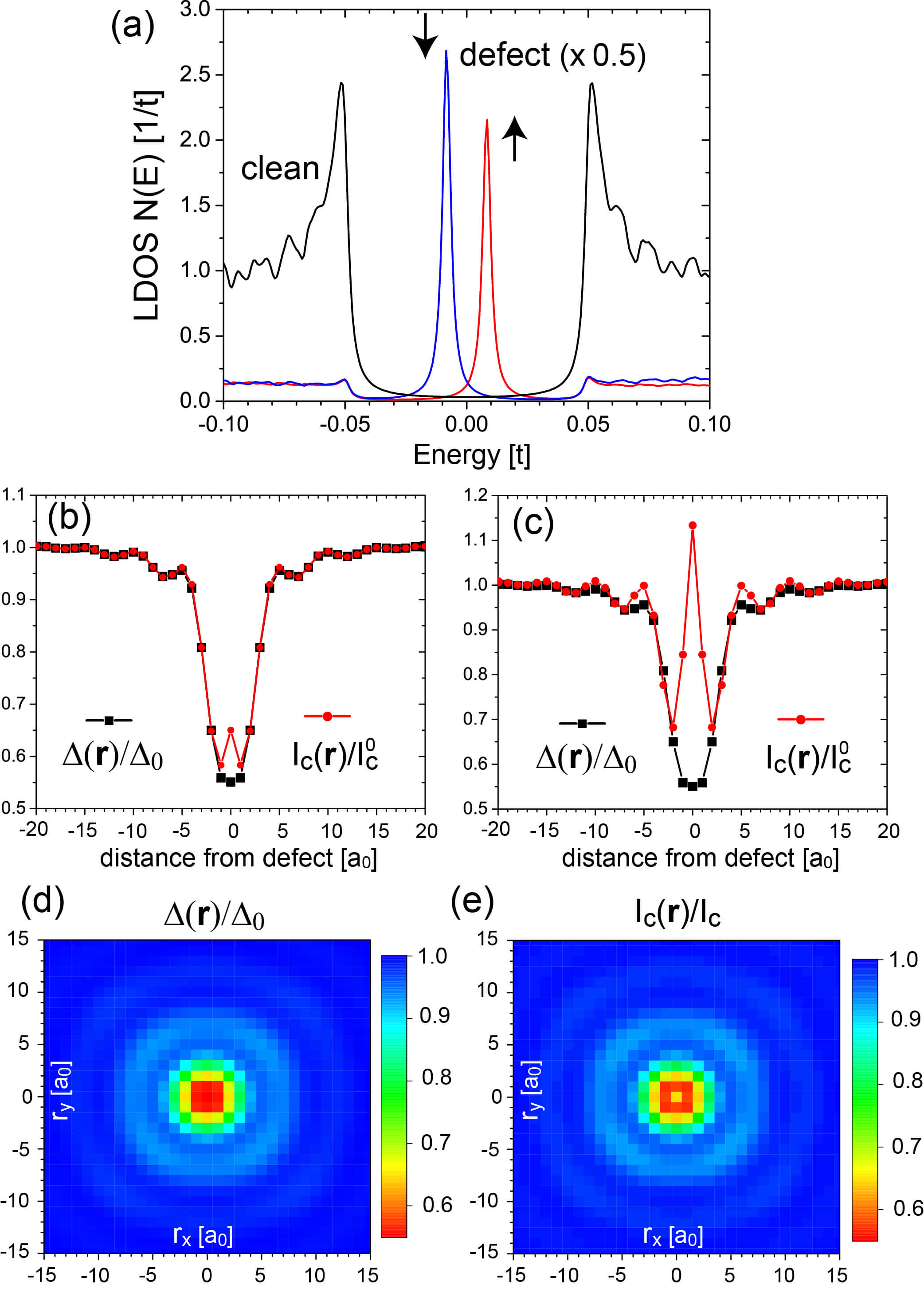}%
 \caption{(a) LDOS at ${\bf r}=(1,0)$ for a magnetic defect with $J_0=2t$ located at ${\bf R}=(0,0)$, and for a clean $s$-wave superconductor with $\Delta_0=0.05t$ and $V_0 = -2.45t$, computed for a $401 \times 401$ system size.  Spatial dependence of the normalized $\Delta({\bf r})$  and $I_c({\bf r})$ along $r_y=0$ for (b) $\Delta_{tip} = 4 \Delta_0$ and (c) $\Delta_{tip} = 0.5 \Delta_0$. Contour plot of the normalized (d) $\Delta({\bf r})$ and (e) $I_c({\bf r})$ for $\Delta_{tip} = 4 \Delta_0$.}
 \label{fig:mag_defect}
 \end{figure}
While only magnetic defects give rise to impurity bound states inside the superconducting gap -- Shiba states \cite{Shi68,Yaz97,Hat15} -- magnetic as well as non-magnetic defects induce spatial oscillations in the superconducting order parameter, $\Delta ({\bf r})$, which, as we show below, can be imaged through the critical Josephson current $I_c({\bf r})$. There exists, however, a qualitative difference in $I_c$ between these two types of defects that reflects the existence of a Shiba states. To demonstrate this, we begin by considering the spatial form of $\Delta ({\bf r})$ and of $I_c({\bf r})$ around a magnetic defect (see SI Sec.~I). In Fig.~\ref{fig:mag_defect}(a), we present the local density of states (LDOS) in the vicinity of the defect together with that in a clean system. As expected, we find that the magnetic defect induces a  Shiba state with spin-resolved particle-like and hole-like branches inside the superconducting gap. In Fig.~\ref{fig:mag_defect}(b), we present the normalized superconducting order parameter, $\Delta({\bf r})/\Delta_0$, and the Josephson current, $I_c({\bf r})/I_c^0$ along $r_y=0$ for $\Delta_{tip} = 4 \Delta_0$. Here, $\Delta_0$ and $I_c^0$ are the superconducting order parameter and the critical Josephson current in a clean system.  We find that the defect-induced oscillations of $\Delta({\bf r})$ are very well spatially imaged by the Josephson current, thus theoretically confirming the assumptions underlying the JSTS experiments by Hamidian {\it et al.}, \cite{Ham15} and Randeria {\it et al.} \cite{Ran16}. To gain analytic insight into the spatial relation between $\Delta({\bf r})$ and $I_c({\bf r})$, we consider the limit of large tip gap $\Delta_{tip} > \omega_D$ where $\omega_D$ is the Debye energy of the $s$-wave superconductor. In this case, we have for the integrand in Eq.(\ref{eq:Ieq}), ${\rm Im}[F_s\, F_t]={\rm Re}F_t{\rm Im}F_s$, and ${\rm Re}F_t(\omega)$ can be approximated by a constant ${\rm Re}{\bar F}_t$ over the energy range where ${\rm Im}F_s$ possesses the largest spectral weight (see SI Sec.~II). Using Eq.(\ref{eq:OP}) we then obtain from Eq.(\ref{eq:Ieq})
 \begin{equation}
 I_c({\bf r}) \sim {\rm Re}{\bar F}_t \int \frac{d\omega}{2\pi}n_F(\omega)\text{Im}[F_s({\bf r},{\bf r},\omega)] \sim  \Delta({\bf r}) \ .
 \end{equation}
 Thus, $I_c({\bf r})/I_c^0 = \Delta({\bf r})/\Delta_0$, and $I_c({\bf r})$ possesses in general the same spatial dependence as $\Delta({\bf r})$. We find, however, that there exists an interesting exception to this result at the site of the magnetic defect, where $I_c({\bf r})$ exhibits a weak peak while $\Delta({\bf r})$ does not [see Fig.~\ref{fig:mag_defect}(b)]. This peak arises from an enhanced tunneling of Cooper pairs from the tip into the Shiba state, whose largest spectral weight resides at the site of the defect (see SI Sec.~II), and thus counteracts the general suppression of $\Delta({\bf r})$ in the vicinity of the defect. As the main contribution to this peak arises from Re$F_t$Im$F_s$ in the integral of Eq.(\ref{eq:Ieq}) (see SI Sec.~II), we expect that the peak height further increases as $\Delta_{tip}$ (and hence the enhancement of Re$F_t$ near $\Delta_{tip}$, see SI Sec.~II) approaches the energy of the Shiba state. This expectation is borne out by our results for a smaller tip gap $\Delta_{tip} = 0.5 \Delta_0$ [see Fig.\ref{fig:mag_defect}(c)], which shows an even stronger enhancement of the peak in $I_c$ near the defect. This peak in $I_c$ is therefore a direct signature of the impurity induced Shiba state and thus absent for non-magnetic defects (see below). The peak's height is not only affected by the value of $\Delta_{tip}$, but also the strength of the magnetic scattering as well as the electronic structure of the superconductor (see SI Sec.~III). However, even in presence of a strong peak in $I_c$ at the defect site, the spatial dependence of $I_c$ in all other regions still reflects that of $\Delta({\bf r})$. This is also confirmed by the spatial plots of the normalized $\Delta({\bf r})$ and $I_c({\bf r})$, shown in Figs.~\ref{fig:mag_defect}(d) and (e), respectively.  Here, the spatially circular oscillations in $\Delta({\bf r})$ and of $I_c({\bf r})$ reflect the form of the underlying circular Fermi surface [see Fig.~\ref{fig:tunneling}(b)], with their wavelength of $\lambda_F/2$ arising from $2k_F$ scattering.

\begin{figure}
\includegraphics[width=8cm]{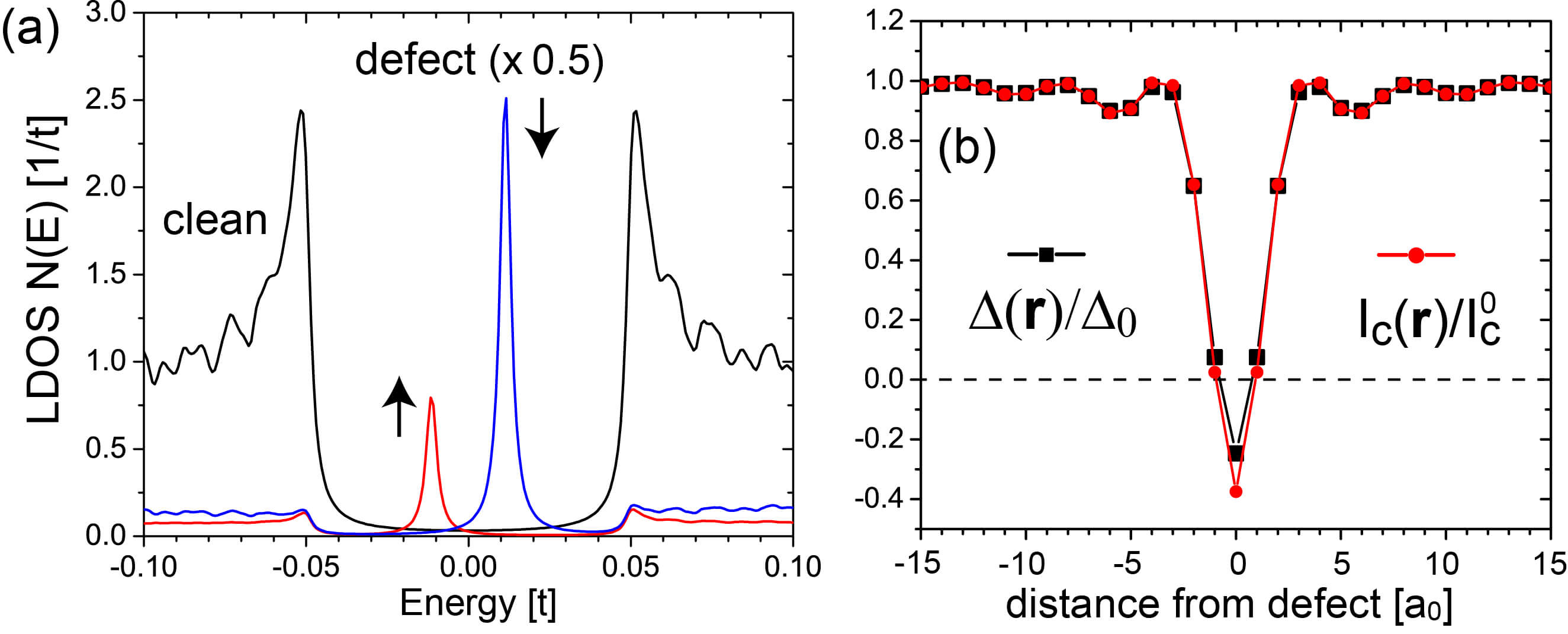}%
 \caption{(a) LDOS at ${\bf r}=(1,0)$ for a magnetic defect with $J_0=2.5t>J_c$ located at ${\bf R}=(0,0)$, and for a clean $s$-wave superconductor.
 (b) Spatial dependence of the normalized $\Delta({\bf r})$  and $I_c({\bf r})$ along $r_y=0$ for $\Delta_{tip} = 4 \Delta_0$}
 \label{fig:PT}
 \end{figure}

As the magnetic scattering strength, $J_0$, is increased and exceeds a critical value, $J_c$, the superconductor undergoes a phase transition in which its ground state changes from a singlet $S=0$ state to a doublet $S=1/2$ state \cite{Sak70,Sal97,Morr06}. Simultaneous with this phase transition, the particle- and hole-like branches of the Shiba state cross at zero energy \cite{Hat15}, and the superconducting order parameter changes sign at the site of the defect. A comparison of the LDOS near a magnetic defect with $J_0=2.5t >J_{c}$ in Fig.~\ref{fig:PT}(a) with the LDOS for $J_0=2.0 t< J_{c}$ in Fig.~\ref{fig:mag_defect}(a) shows that the two branches of the Shiba state have crossed zero energy, as the particle-like (hole-like) branches for $J_0< J_{c}$ and $J_0> J_{c}$ possess different spin character.
Moreover, the sign of the superconducting order parameter changes at the site of the defect [see Fig.~\ref{fig:PT}(b)] which is mirrored by a sign change in the Josephson current. This sign change in $I_c$ as a function of distance from the defect is a direct signature of the $S=1/2$ ground state of the superconductor. Thus, we have demonstrated that the spatial form of the Josephson current not only reflects that of $\Delta({\bf r})$, but that it is also a probe for the spin ground states of the superconductor, and hence can be employed to detect a quantum phase transition of the system. This opens up the possibility to investigate more complex ground states with even larger spin polarizations, as arise, for example, from quantum interference effects in multi-defect systems \cite{Morr06}.

\begin{figure}
 \includegraphics[width=8cm]{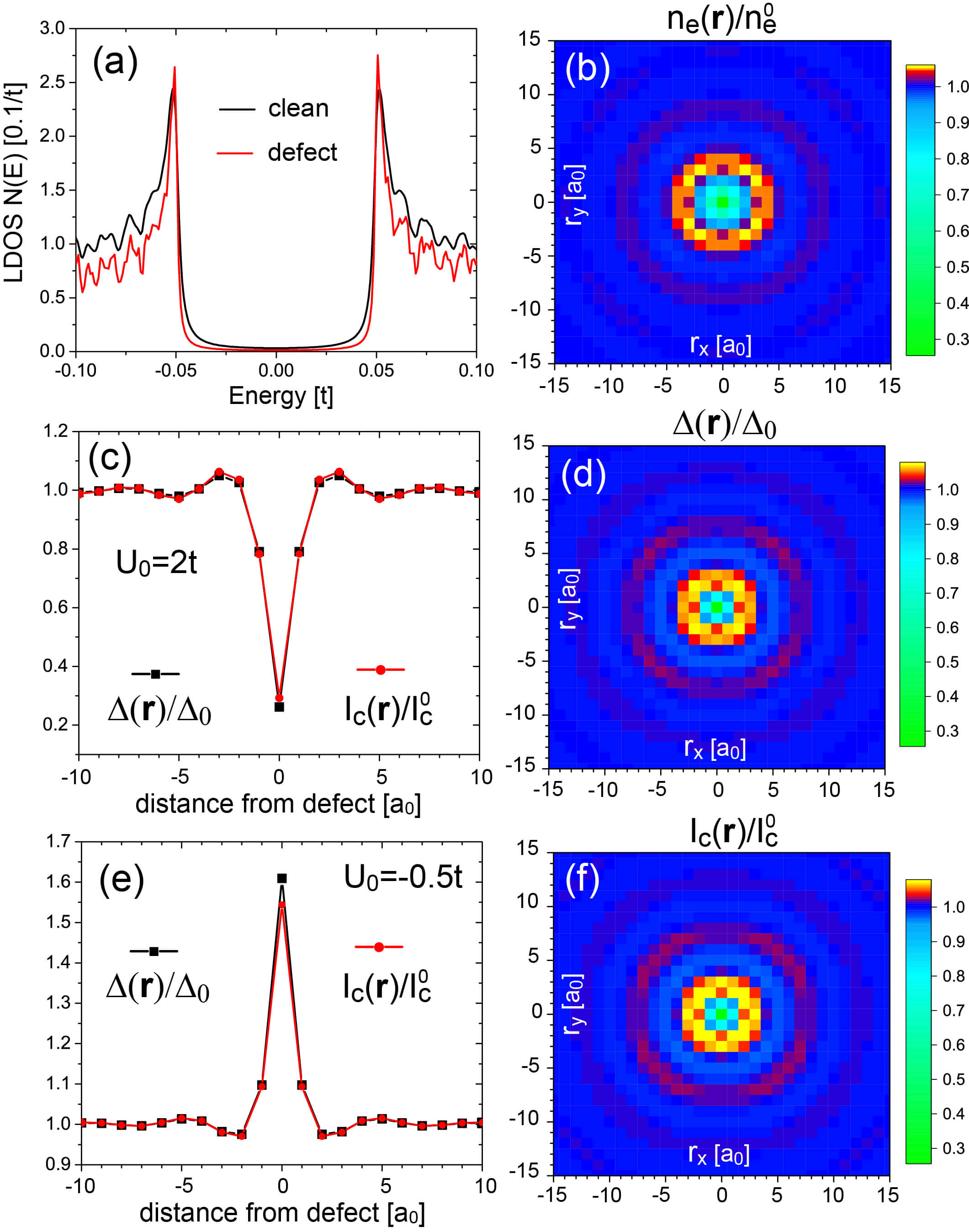}%
 \caption{(a) LDOS at ${\bf r}=(1,0)$ in an $s$-wave superconductors with a non-magnetic defect located at ${\bf R}=(0,0)$ with repulsive scattering potential $U_0=2t$. Contour plot of the normalized (b) charge density $n_e({\bf r})$, (d) $\Delta({\bf r})$, and (f) $I_c({\bf r})$. Spatial cut along $r_y=0$ of the normalized $\Delta({\bf r})$ and $I_c({\bf r})$ for (c) $U_0=2t$, and (e) $U_0=-0.5t$, and $\Delta_{tip}=4 \Delta_0$.}
 \label{fig:npot_defect}
 \end{figure}

 In contrast to magnetic defects, non-magnetic (potential) defects do not induce impurity states inside the superconducting gap \cite{And59}, as follows from a plot of the LDOS near the site of a repulsive potential defect with $U_0=2t$ in Fig.~\ref{fig:npot_defect}(a). However, the scattering off non-magnetic defects induces oscillations in the electron charge density, $n_e$, [Fig.~\ref{fig:npot_defect}(b)] which in turn give rise to spatial variations of the superconducting order parameter [Figs.~\ref{fig:npot_defect}(c) and (d)].
While magnetic defects lead to an overall suppression of the superconducting order parameter, non-magnetic defects, through oscillations in $n_e({\bf r})$, give rise to spatial regions in which $\Delta({\bf r})$ is enhanced or suppressed. These spatial oscillations can again be imaged by the Josephson current, as demonstrated by the spatial contour plots of $\Delta({\bf r})$ and $I_c({\bf r})$ in Fig.~\ref{fig:npot_defect}(d) and (f), respectively, and the line cut in Fig.~\ref{fig:npot_defect}(c).
Due to the absence of a Shiba state, the tunneling of Cooper pairs from the tip into the superconductor is not enhanced at the site of the defect, and no peak in $I_c$ is therefore found. Moreover, a non-magnetic defect with an attractive scattering potential, $U_0=-0.5t$, leads to an enhancement of the charge density and hence the superconducting order parameter near the defect that is also reflected in the spatial form of $I_c({\bf r})$ [see Fig.~\ref{fig:npot_defect}(e)]. Thus, the enhancement or decrease of the critical current in the vicinity of a non-magnetic defect can distinguish between its attractive and repulsive scattering potential.

\begin{figure}
\includegraphics[width=8cm]{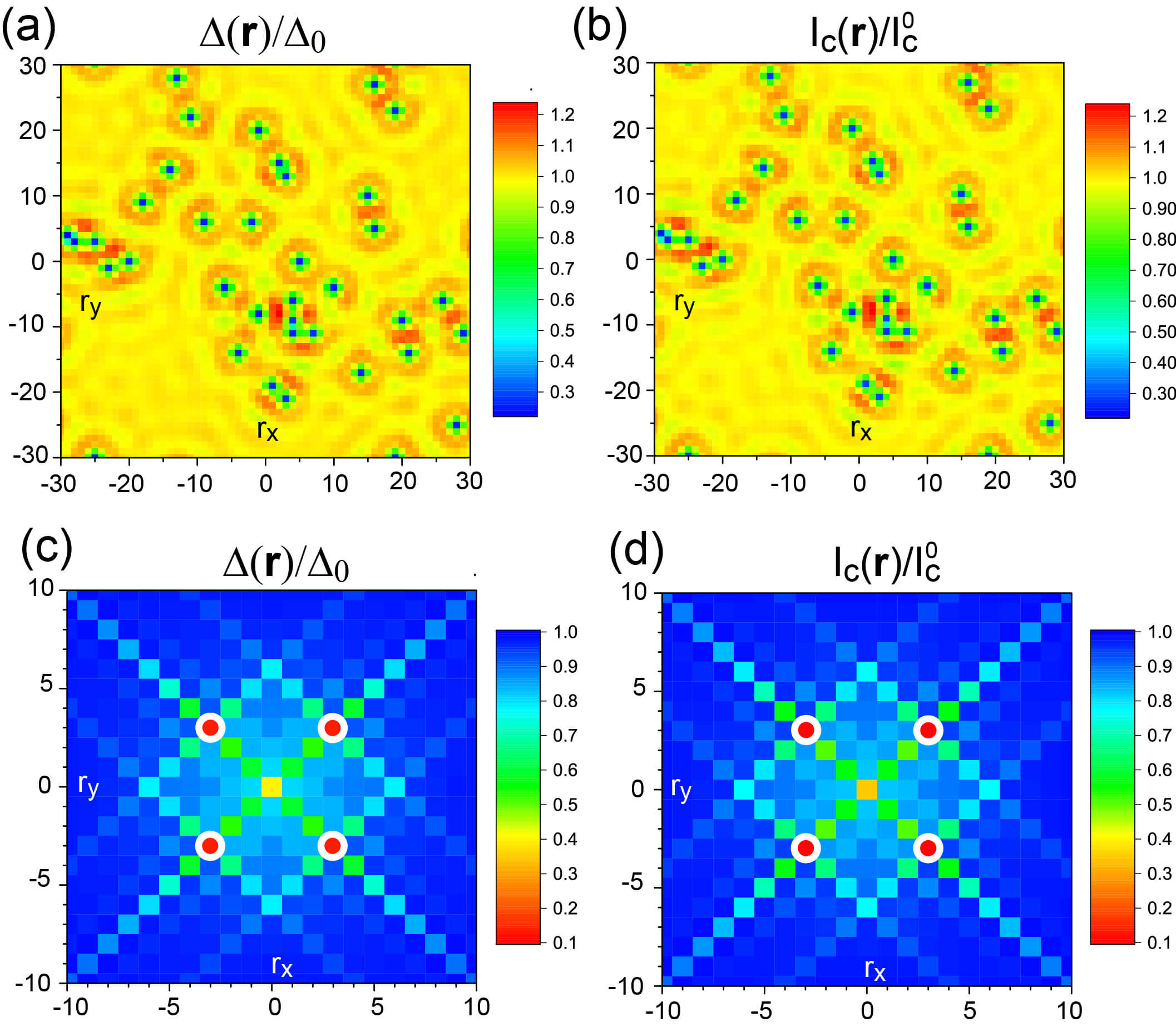}%
 \caption{(a) $\Delta({\bf r})/\Delta_0$ and (b) $I_c({\bf r})/I_c^0$ for a concentration of 1\% randomly distributed non-magnetic defects with $U_0=2t$.
(c) $\Delta({\bf r})/\Delta_0$ and (d) $I_c({\bf r})/I_c^0$ in the presence of four magnetic defects (whose locations are indicated by open white circles) with $J_0=t$, $\Delta_{tip}=4 \Delta_0$ and the dashed blue Fermi surface in Fig.~\ref{fig:tunneling}(b) obtained with $\mu=0$.}
 \label{fig:4def}
 \end{figure}

The ability to image spatial oscillations of the superconducting order parameter via $I_c$ are independent of the particular form of the material's Fermi surface or the strength of the scattering potential (see SI Sec.~III). Moreover, $\Delta({\bf r})$ can not only be mapped around isolated defects, as discussed above, but also in disordered superconductors with a random distribution of defects \cite{Lang02}, as shown in Figs.~\ref{fig:4def}(a) and (b). Here, we present $\Delta({\bf r})/\Delta_0$ and of $I_c({\bf r})/I_c^0$, respectively, for a concentration of 1\% randomly distributed non-magnetic defects with $U_0=2t$.
While the interference of electrons scattered by multiple defects can lead to spatial regions in which the superconducting order parameter is significantly enhanced or suppressed, the spatial form of $I_c({\bf r})$ again very well images that of $\Delta({\bf r})$. The critical Josephson current can even be employed to image  "virtual defects", i.e., regions in which the superconducting order parameter is strongly suppressed without the existence of defects. Such virtual defects can be created using quantum interference effects, as shown in Fig.~\ref{fig:4def}(c), where we present the superconducting order parameter in the presence of four defects located at sites denoted by white open circles and $\mu=0$, yielding the dashed blue Fermi surface in Fig.~\ref{fig:tunneling}(b). Interference effects give rise to an additional strong suppression of $\Delta({\bf r})$ in the center of the superconductor -- the virtual defect -- which is again captured by $I_c({\bf r})$, as shown in Fig.~\ref{fig:4def}(d).

The results discussed above demonstrate that the spatial dependence of the critical Josephson current images that of the superconducting order parameter. This provides the missing crucial link for interpreting the JSTS experiments by Hamidian {\it et al.} \cite{Ham15} and Randeria {\it et al.} \cite{Ran16}. The ability to image $\Delta({\bf r})$ in an $s$-wave superconductor raises the question of whether JSTS can also be used to investigate the order parameter of unconventional superconductors, such as the cuprate, iron-based or heavy fermion superconductors. The spatially extended nature of the superconducting order parameter in these materials will require the use of more extended JSTS tips. Work is currently under way to investigate this interesting question.

\begin{acknowledgments}
We would like to thank C. Ast and M. Hamidian for helpful discussions.  This work was supported by the U. S. Department of Energy, Office of Science, Basic Energy Sciences, under Award No. DE-FG02-05ER46225.
\end{acknowledgments}

\end{document}


\fontsize{11}{13}

\begin{center}
{\large {\bf  Supplemental Online Information for}} \\[0.5cm]
\end{center}

\title{Imaging the Spatial Form of a Superconducting Order Parameter via Josephson Scanning Tunneling Spectroscopy}

\author{Martin Graham and  Dirk K. Morr}

\affiliation{Department of Physics, University of Illinois at Chicago, Chicago,
IL 60607}

\date{\today}

\maketitle

\section{Theoretical Methods}
\label{sec:method}

\subsection{Definition of Green's functions in real space}
To compute the spatial dependence of the local $s$-wave order parameter, $\Delta({\bf r})$, as well as the critical Josephson current, $I_c({\bf r})$, in the presence of defects, we rewrite the Hamiltonian in Eq.(1) of the main text in matrix form introducing the spinor
\begin{align}
\Psi^\dagger = \left(c^\dagger_{1,\uparrow}, c_{1,\downarrow}, \ldots, c^\dagger_{i,\uparrow}, c_{i,\downarrow}, \ldots, c^\dagger_{N,\uparrow}, c_{N,\downarrow} \right)
\label{eq:psi}
\end{align}
where $N$ is the number of sites in the $s$-wave superconductor, and $i=1,...,N$ is the index for a site ${\bf r}$ in the system. The Hamiltonian in Eq.(1) of the main text can then be written as
\begin{align}
H_s = \Psi^\dagger {\hat H}_s \Psi \ .
\end{align}
We next define a retarded Green's function matrix of the system via
\begin{align}
{\hat G}_s (\omega+i \delta) = \left[ (\omega+i \delta) {\hat 1} - {\hat H}_s \right]^{-1}
\end{align}
where ${\hat 1}$ is the $(N \times N)$ identity matrix and $\delta = 0^+$. The local anomalous Green's function at site ${\bf r}$ (with index $i$), $F_s({\bf r},{\bf r},\omega)$, is then the  given by the $(2i-1,2i)$ element of ${\hat G}_s$. Moreover, we take the anomalous Green's function of the tip to be that of a bulk system given by
\begin{align}
F_t (\omega) = - N_0 \Delta_{tip}  \frac{ \pi i }{\sqrt{(\omega+i \delta)^2 - \Delta^2_{tip}}} {\rm sgn}(\omega)
\label{eq:AGF_tip}
\end{align}
where $N_0$ is the density of states in the tip. This form implies that $F_t$ possess a non-zero real part only for $|\omega|<\Delta_{tip}$, and a non-zero imaginary part only for $|\omega|>\Delta_{tip}$.

\subsection{$I_c$ for a magnetic defect}

A magnetic defect breaks the time-reversal symmetry of the system, implying that the spin-$\uparrow$ and spin-$\downarrow$ contributions to the Josephson current are different. Given the definition of the spinor in Eq.(\ref{eq:psi}), the expression for the critical Josephson current in Eq.~(7) of the main text is that of the spin-$\uparrow$ contribution, i.e., $I_{c,\uparrow}(J)$, where $J$ is the scattering strength of the magnetic defect. Note that $J \rightarrow -J$ implies that the direction of the defects' magnetic moment is flipped with respect to the spin quantization axis of the $s$-wave superconductor and the tip. To obtain the total Josephson current in the presence of a magnetic defect, $I_c(J) = I_{c,\uparrow}(J) + I_{c,\downarrow}(J)$, we use the identity
\begin{align}
I_{c,\uparrow}(J)  &= I_{c,\downarrow}(-J) \nonumber \\
I_{c,\downarrow}(J)  &= I_{c,\uparrow}(-J)
\end{align}
As a result,
\begin{align}
I_c(J) = I_{c,\uparrow}(J) + I_{c,\downarrow}(J) = I_{c,\uparrow}(J) + I_{c,\uparrow}(-J)
\end{align}

\section{Enhanced Josephson current due to tunneling into Shiba states}

\begin{figure}
\includegraphics[width=15cm]{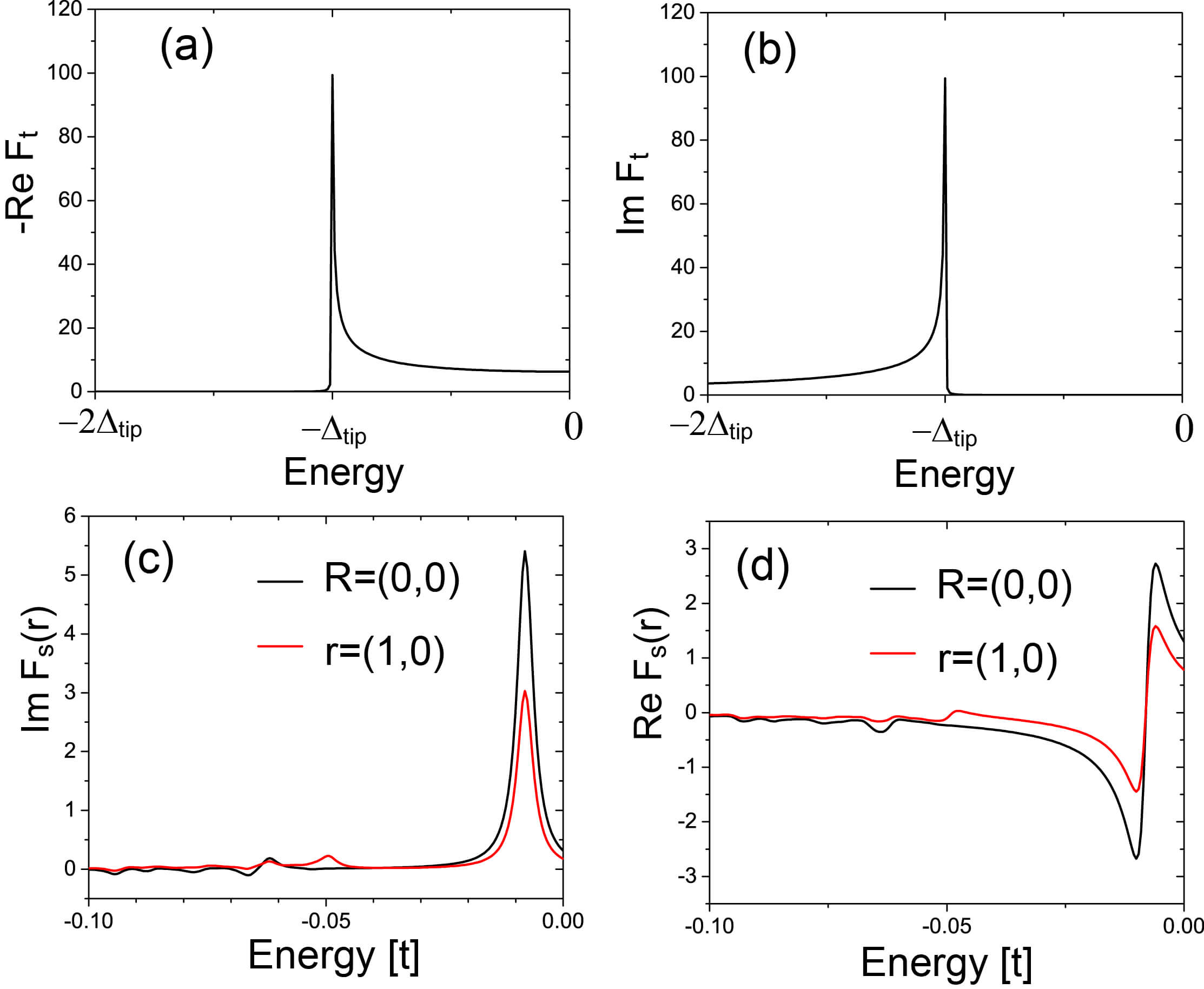}%
 \caption{Energy dependence of the anomalous Green's function, $F_t$, of the tip [see Eq.(\ref{eq:AGF_tip})]: (a) $-{\rm Re}F_t$, and  (b) ${\rm Im}F_t$. Energy dependence of the anomalous Green's function, $F_s$, of the $s$-wave superconductor at the site of a magnetic defect ${\bf R}=(0,0)$ with $J=2t$ and $\mu=-3.618t$, and its nearest neighbor site ${\bf r}=(1,0)$ (corresponding to the case shown in Fig.~2 of the main text): (c) ${\rm Im}F_s$, and  (d) ${\rm Re}F_s$.}
 \label{fig:Shiba}
 \end{figure}

In Fig.~2(b) and 2(c) of the main text, we showed that the critical Josephson current, $I_c({\bf r})$, can exhibit a peak at the site of a magnetic defect due to enhanced Josephson tunneling of Cooper pairs into the defect-induced Shiba state. To demonstrate the origin of this enhancement in more detail, we plot in Fig.~\ref{fig:Shiba} the anomalous Green's function of the tip, $F_t$ [Figs.~\ref{fig:Shiba}(a) and (b)], and of the $s$-wave superconductor, $F_s$ [Figs.~\ref{fig:Shiba}(c) and (d)]. The Shiba state possesses a hole- and particle-like branch at $\pm \omega_b$.
The critical Josephson current [see Eq.(7) of the main text] is given by the frequency integral over ${\rm Im}[F_tF_s] = {\rm Re}F_t {\rm Im}F_s + {\rm Im}F_t {\rm Re}F_s $  in the range (at zero temperature) $\infty < \omega <0$. The factors of the first term, ${\rm Re}F_t$ and $ {\rm Im}F_s$, are shown in Figs.~\ref{fig:Shiba}(a) and (c), while the factors of the second term, ${\rm Im}F_t$ and $ {\rm Re}F_s$, are shown in Figs.~\ref{fig:Shiba}(b) and (d). As ${\rm Im}F_t =0$ for $|\omega|<\Delta_{tip}$, it is clear from these plots that for $\Delta_{tip}>\omega_b$, the main contribution to $I_c$ arises from ${\rm Re}F_t {\rm Im}F_s$. Moreover, as $\Delta_{tip}>\Delta_0$ is reduced and approaches the energy of the Shiba state $\omega_b$, the overlap between the peak in ${\rm Im}F_s$ representing the Shiba state, and the enhanced value of ${\rm Re}F_t$ near $\Delta_{tip}$ is increased, leading to a larger peak in $I_c$ at the defect site. This result is generic and independent of the particular form of $F_t$, as the superconducting gap in the tip leads to an onset of non-zero ${\rm Im}F_t$ at $|\omega|=\Delta_{tip}$ thus requiring (by Kramers-Kronig transformation) ${\rm Re}F_t$ to increase in magnitude near $\Delta_{tip}$. As the spectral weight of the Shiba state away form the defect site [see ${\bf r}=(1,0)$ in Fig.~\ref{fig:Shiba}(c)] is smaller than that at the defect site, the enhancement in $I_c$ due to Cooper pair tunneling into the Shiba state is more pronounced at the defect site. This explains the spatial form of $I_c({\bf r})$ and its evolution with $\Delta_{tip}$ shown in Fig.~2 of the main text, and in Fig.~\ref{fig:mag_FS3}.

\section{Superconducting Order Parameter and critical Josephson current}

\begin{figure}
\includegraphics[width=15cm]{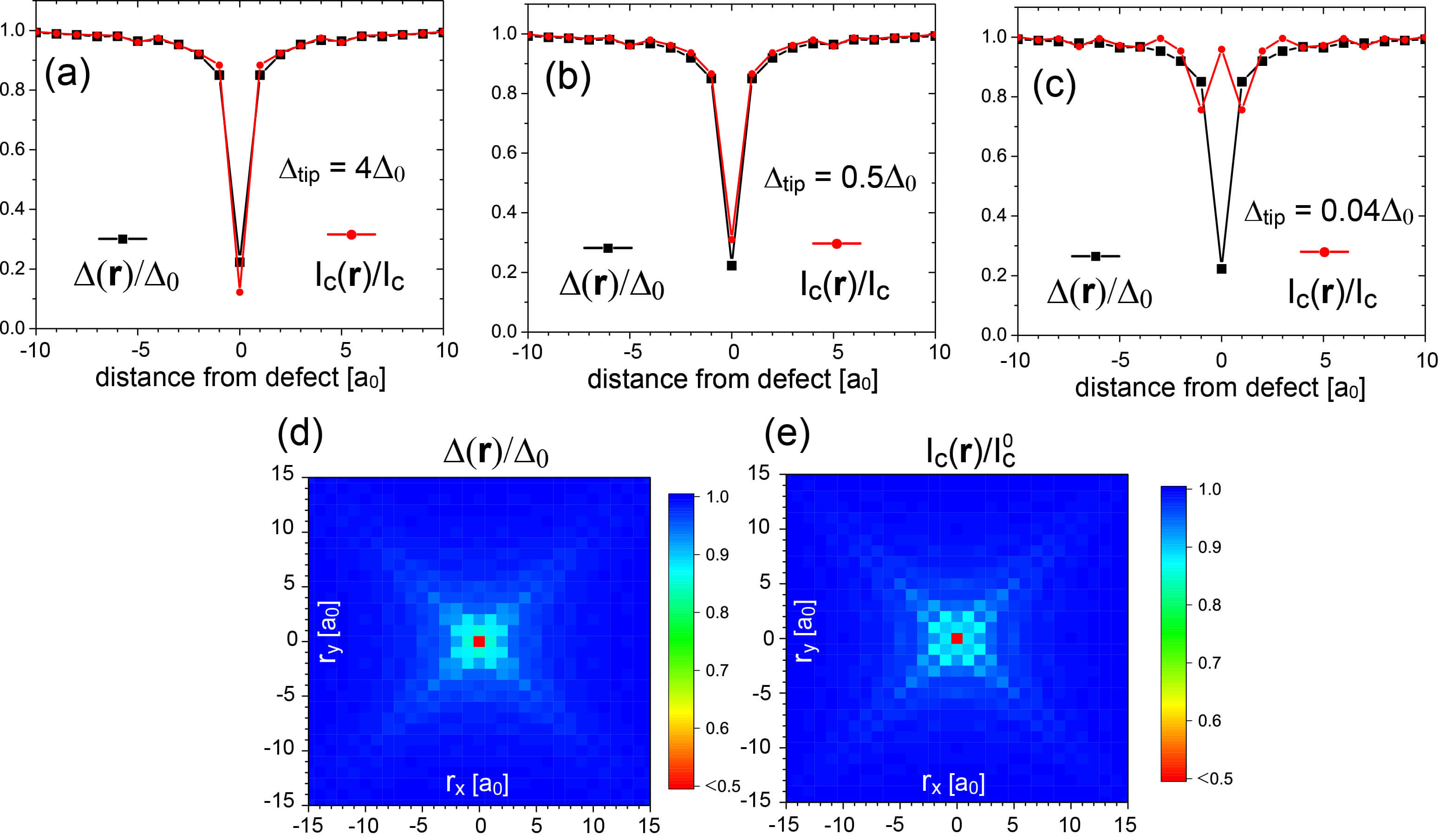}%
 \caption{Spatial dependence of the normalized $\Delta({\bf r})$  and $I_c({\bf r})$ along $r_y=0$ for a magnetic defect with $J=2.0t$ and $\mu=t$ (yielding the red Fermi surface shown in Fig.~1(b) of the main text) for (a) $\Delta_{tip} = 4 \Delta_0$, (b) $\Delta_{tip} = 0.5 \Delta_0$, and (c) $\Delta_{tip} = 0.04 \Delta_0$. Contour plot of the normalized (d) $\Delta({\bf r})$  and (e) $I_c({\bf r})$  for $\Delta_{tip} = 4 \Delta_0$.}
 \label{fig:mag_FS3}
 \end{figure}

In this section, we provide further examples that demonstrate the ability to image the spatial form of $\Delta({\bf r})$ through local measurements of $I_c({\bf r})$ in the vicinity of defects in an $s$-wave superconductor. In Figs.~\ref{fig:mag_FS3} (a) and (b) we present a line cut of the normalized $\Delta({\bf r})/\Delta_0$, and  $I_c({\bf r})/I_c^0$ for an $s$-wave superconductor with $\mu=t$ (yielding the red Fermi surface shown in Fig.~1(b) of the main text) and a magnetic defect with $J_0=2t$, using the same ratios of $\Delta_{tip}/\Delta_0$ as in Figs.~2(b) and (c) of the main text. We again find that the spatial form of the superconducting order parameter is well reproduced by that of $I_c({\bf r})$, as also follows from a contour plot of $\Delta({\bf r})$ and $I_c({\bf r})$ (for $\Delta_{tip} = 4 \Delta_0$) shown in Figs.~\ref{fig:mag_FS3} (d) and (e).
 However, in contrast to the results shown in Figs.~2(b) and (c) of the main text, $I_c({\bf r})$ does not exhibit a peak at the defect site. This "missing" peak is a direct consequence of the rapid suppression of the superconducting order parameter in the immediate vicinity of the defect (this rapid suppression differs from that shown in Fig.~2(b) of the main text, and is a result of the different electronic structures, resulting from changes in the chemical potential). For the values of $\Delta_{tip}$ used in Figs.~\ref{fig:mag_FS3} (a) and (b), the enhanced tunneling into the Shiba state cannot overcome the rapid suppression of $\Delta({\bf r})$ [and hence $F_s({\bf r},{\bf r}, \omega)$] in the vicinity of the defect [though the relative strength of $I_c({\bf r})$ at the defect site is increased when $\Delta_{tip}$ is reduced from $4 \Delta_0$ in Fig.~\ref{fig:mag_FS3}(a) to $0.5 \Delta_0$ in Fig.~\ref{fig:mag_FS3}(b)]. Only when $\Delta_{tip}$ is further reduced to $\Delta_{tip} = 0.04 \Delta_0$ and thus brought closer to the bound state energy of the Shiba state, $\omega_b=0.0015t=0.03\Delta_0$, is the enhanced tunneling sufficiently strong to result in a peak in $I_c({\bf r})$ at the defect site [see Fig.~\ref{fig:mag_FS3}(c)].

\begin{figure}
\includegraphics[width=15cm]{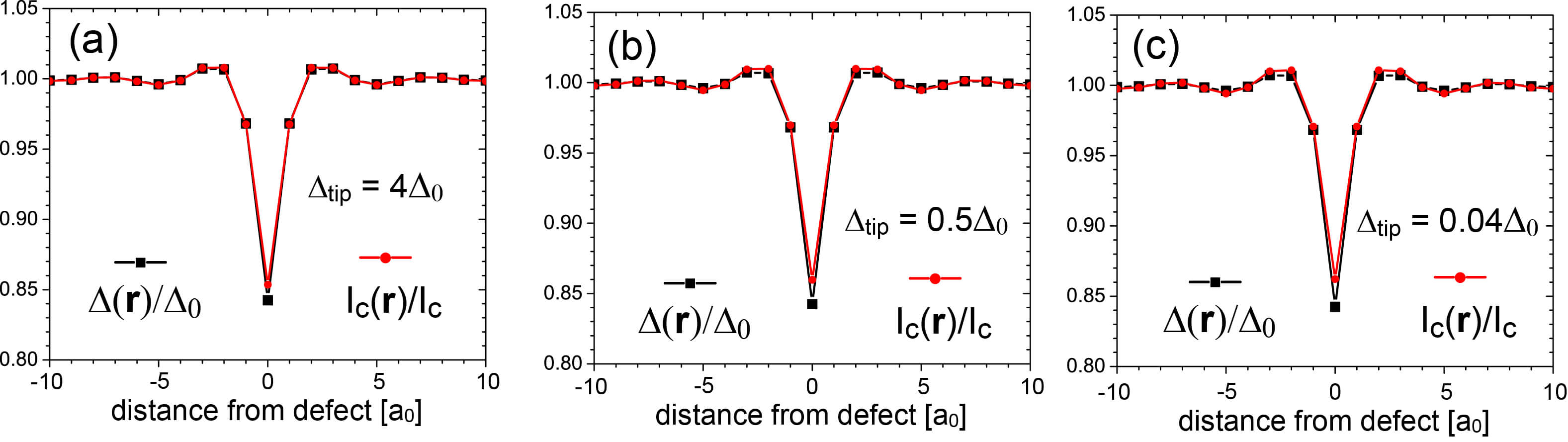}%
 \caption{Spatial dependence of the normalized $\Delta({\bf r})$  and $I_c({\bf r})$ along $r_y=0$ for a weakly scattering non-magnetic defect with $U_0=0.2t$ and $\mu=-3.618t$ and (a) $\Delta_{tip} = 4 \Delta_0$, (b) $\Delta_{tip} = 0.5 \Delta_0$, and (c) $\Delta_{tip} = 0.04 \Delta_0$. }
 \label{fig:pot_FS1}
 \end{figure}

The ability to spatially image the superconducting order parameter is independent of the scattering strength of the defect. To demonstrate this, we consider a weakly scattering non-magnetic defect with $U_0=0.2t$ in a superconductor with $\mu=-3.618$ [yielding the black Fermi surface shown in Fig.~1(b) of the main text].  The line cuts of the normalized $\Delta({\bf r})/\Delta_0$, and  $I_c({\bf r})/I_c^0$ presented in Fig.~\ref{fig:pot_FS1} show again a very similar spatial dependence. Note that in contrast to the above case of a magnetic defect, the normalized Josephson current $I_c({\bf r})/I_c^0$ at the site of the defect remains approximately constant with decreasing $\Delta_{tip}$, due to the absence of a Shiba state [here, the same ratios of $\Delta_{tip}/\Delta_0$ as in Figs.~\ref{fig:mag_FS3}(a)-(c) were used].
